\documentclass[prb,,superscriptaddress,twocolumn,showpacs,preprintnumbers,amsmath,amssymb,reprint]{revtex4-1}

\usepackage{graphicx}
\usepackage{amsmath}
\usepackage{amssymb}
\usepackage{color}
\usepackage{dcolumn}
\usepackage{epsfig}
\usepackage{bm}
\usepackage{subfigure}
\usepackage[urlcolor=blue]{hyperref}
\hypersetup{backref, colorlinks=true, linkcolor=blue, citecolor=blue}

\begin{document}

\title{Origin of efficient thermoelectric performance in half-Heusler FeNb$_{0.8}$Ti$_{0.2}$Sb}

\author{Hong-Jie Pang}
\affiliation{Center for High Pressure Science and Technology Advanced Research, Shanghai 201203, China}

\author{Chen-Guang Fu}
\affiliation{State Key Laboratory of Silicon Materials, School of Materials Science and Engineering, Zhejiang University, Hangzhou 310027, China}

\author{Hao Yu}
\affiliation{Center for High Pressure Science and Technology Advanced Research, Shanghai 201203, China}

\author{Liu-Cheng Chen}
\affiliation{Center for High Pressure Science and Technology Advanced Research, Shanghai 201203, China}

\author{Tie-Jun Zhu}
\affiliation{State Key Laboratory of Silicon Materials, School of Materials Science and Engineering, Zhejiang University, Hangzhou 310027, China}

\author{Xiao-Jia Chen}
\email{xjchen@hpstar.ac.cn}
\affiliation{Center for High Pressure Science and Technology Advanced Research, Shanghai 201203, China}

\date{\today}

\begin{abstract}
A half-Heusler material FeNb$_{0.8}$Ti$_{0.2}$Sb has been identified as a promising thermoelectric material due to its excellent thermoelectric performance at high temperatures. The origins of the efficient thermoelectric performance are investigated through a series of low-temperature (2 - 400 K) measurements. The high data coherence of the low and high temperatures is observed. An optimal and nearly temperature-independent carrier concentration is identified, which is ideal for the power factor. The obtained single type of hole carrier is also beneficial to the large Seebeck coefficient. The electronic thermal conductivity is found to be comparable to the lattice thermal conductivity and becomes the dominant component above 200 K. These findings again indicate that electron scattering plays a key role in the electrical and thermal transport properties. The dimensionless figure of merit is thus mainly governed by the electronic properties. These effects obtained at low temperatures with the avoidance of possible thermal fluctuations together offer the physical origin for the excellent thermoelectric performance in this material.
\end{abstract}

\pacs{65.40.-b, 72.20.Pa}

\maketitle

\section{INTRODUCTION}

Thermoelectric materials have attracted a great deal of interest due to their remarkable applications in meeting the world's demand for generating electricity from waste heat and solid-state Peltier coolers\,\cite{dis, nol1, sny}. The thermoelectric efficiency of a material is determined by the dimensionless figure of merit\,\cite{trit1, mah}, defined as $zT=S^2\sigma T/\kappa$=PF $T/\kappa$, where $S$ is the Seebeck coefficient, $\sigma$ is the electrical conductivity, $T$ is the absolute temperature, and $\kappa$ is the total thermal conductivity (including the lattice contribution $\kappa_l$ and the electron contribution $\kappa_e$), and PF is the power factor (PF=$S^{2}\sigma$). $zT$=3 is needed for thermoelectric energy converters to complete with mechanical power generation and active refrigeration. However, state-of-the-art commercially available thermoelectric materials have a peak $zT$ value less than unity. As a result, a material suitable for thermoelectric applications must be optimized through electrical conductivity, Seebeck coefficient and thermal conductivity. However, aside from the independent parameter lattice thermal conductivity, the other transport properties (electrical conductivity, Seebeck coefficient, and electronic thermal conductivity) cannot be independently tuned in an effort to increase $zT$. Because the properties are interdependent via the carrier concentration ($n$) in a given thermoelectric material\,\cite{sny,nol2,nol3}. Therefore, the main conventional efforts for maximizing $zT$ of thermoelectric materials are carrier concentration optimization\,\cite{sny} and lattice thermal conductivity reduction\,\cite{sale,hsu,poudel}. It is well known that the optimal carrier concentration depends on temperature and the band structure of thermoelectric semiconductors. As a consequence, there are two major approaches that are pursued separately or in conjunction to achieve higher $zT$: One is to find new crystalline materials with unique structure property relationships that yield the desired combination of properties\,\cite{sny,trit2,cyu,kau,bro}, and the other is to utilize band engineering\,\cite{her1,her2}, alloying\,\cite{bis} or nanostructuring\,\cite{yang1,sak} to tune electrical and thermal transport properties.

Half-Heusler compounds with a valence electron count of 18 have recently been identified as promising thermoelectric materials due to the unique XYZ structures\,\cite{cgf,tjz3,jhe}. These phases are well-known semiconductors with a narrow energy gap and sharp slope of density of states near the Fermi level, which could potentially provide a higher Seebeck coefficient and moderate electrical conductivity\,\cite{ali,gal,yang2,sim}. Nevertheless, the lattice thermal conductivity is relatively high\,\cite{hon,uhe,xia,sek}. Among them, it's noteworthy that $p$-type FeNb$_{0.8}$Ti$_{0.2}$Sb is more competitive not only because the elements are inexpensive and Hf-free but also due to it possessing a relatively low lattice thermal conductivity. More importantly, FeNb$_{0.8}$Ti$_{0.2}$Sb exhibits excellent thermoelectric performance at high temperatures ($>$ 900 K). The $zT$ is superior to the optimized typical half-Heusler compounds\,\cite{tjz1,dow,che}, and especially, the maximum $zT$ (1.1 at 1100 K)\,\cite{fu} is almost twice as high as that of the most widely used $p$-type silicon-germanium thermoelectric materials\,\cite{poudel,zeb,yu,jos,poo}. Fu \emph{et al}.\,\cite{fu} have also confirmed the good experimental repeatability and high-temperature stability of FeNb$_{0.8}$Ti$_{0.2}$Sb. Although the excellent high-temperature thermoelectric performance of FeNb$_{0.8}$Ti$_{0.2}$Sb has been presented, the physical mechanisms are not clear\,\cite{Ran}. The study of material properties at low temperatures without thermal fluctuations is essential to have a real understanding of physical origins of its good performance at high temperatures.

In this work, we present a series of low-temperature investigation of FeNb$_{0.8}$Ti$_{0.2}$Sb in order to obtain the physical origins of its excellent thermoelectric performance at high temperatures. The physical mechanisms for low-temperature electrical and thermal properties are revealed. Moreover, the high data coherence of low and high temperature is observed. Thus, the physical mechanisms at low temperatures are extended to high temperatures.

\section{EXPERIMENTAL DETAILS}

The sample ingot with nominal composition FeNb$_{0.8}$Ti$_{0.2}$Sb used in the experiment was synthesized by levitation melting\,\cite{fu}. The obtained ingot was mechanically milled to obtained fine-grained powders. Afterwards, the powders were immediately  and compacted by using spark plasma sintering at 1123 K for 10 minutes under 65 MPa in a vacuum, for a more detailed explanation refer to Fu \emph{et al}.\,\cite{fu}. The as-sintered samples were annealed at 1123 K for 8 days. Phase structures of the sample were investigated by X-ray diffraction on a RigakuD /MAX-2550PC diffractometer using Cu-K{$\alpha$} radiation ($\lambda$$_{0}$=1.5406 {\AA}) and the chemical compositions were checked by Energy Dispersive Spectrometer on an OXFORD X-Max$^{N}$. Magnetic susceptibility measurements were carried out in the temperature range of 1.8 - 300 K and in the magnetic fields up to 5 T using a Magnetic Property Measurement System (Quantum Design). The electrical conductivity and the Seebeck coefficient as well as the thermal conductivity measurements were performed from 2 K to 400 K by the thermal transport option (TTO) of Physical Property Measurement System (Quantum Design). The Hall coefficient and specific heat measurements were completed in the temperature range of 1.8 - 400 K also using a Physical Property Measurement System (Quantum Design). For high-temperature (300 - 1100 K), the electrical conductivity and Seebeck coefficient were measured on a commercial Linseis LSR-3 system and the thermal conductivity was estimated by a laser flash method on Netzsch LFA457 instrument with a Pyroceram standard.

\section{RESULTS AND DISCUSSION}

\subsection{Structural characterization}

\begin{figure}[tbp]
\centerline{\includegraphics[width=0.48\textwidth]{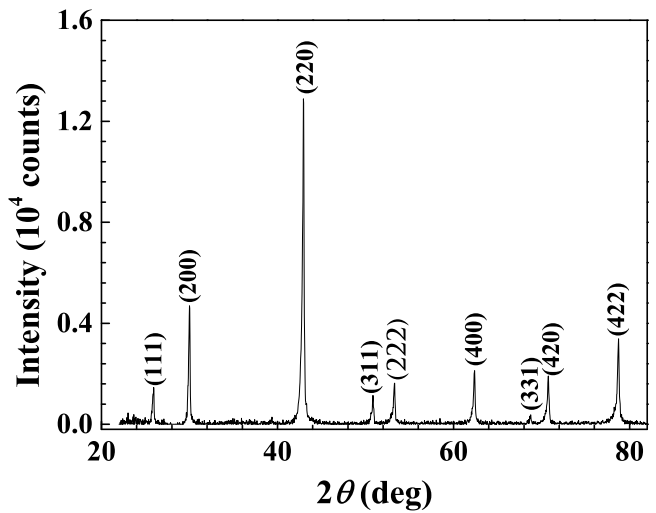}}
\label{modes} \caption{(Color online) X-ray diffraction pattern of FeNb$_{0.8}$Ti$_{0.2}$Sb.}
\end{figure}

\begin{table}[b]
\centering
\caption[modes]{Atomic distribution of FeNb$_{0.8}$Ti$_{0.2}$Sb.}
\begin{tabular*}{8cm}{@{\extracolsep{\fill}}lcccc}
\hline
\textbf{Element} & \textbf{Ti}  & \textbf{Fe}  & \textbf{Nb}  & \textbf{Sb} \\\hline
\vspace{0.05cm}
\textbf{Atomic\%} & 6.31  & 33.09  & 27.29  & 33.32 \\\hline
\end{tabular*}
\end{table}

The X-ray diffraction pattern of the sample, as shown in Fig. 1, was fully indexed within cubic face-centered unit cell with lattice parameter $\emph{a}$=5.951 {\AA}. Compared with the data base, the intensities of the diffraction peaks belong to a space group of $\emph{F}\bar{4}3\emph{m}$ which is consistent with the literature\,\cite{cas}. Table I shows the atomic distribution of the sample. The chemical composition of the sample which was determined from Energy Dispersive Spectrometer is FeNb$_{0.8}$Ti$_{0.2}$Sb.

\subsection{Magnetic characterization}

\begin{figure}[tbp]
\centerline{\includegraphics[width=0.48\textwidth]{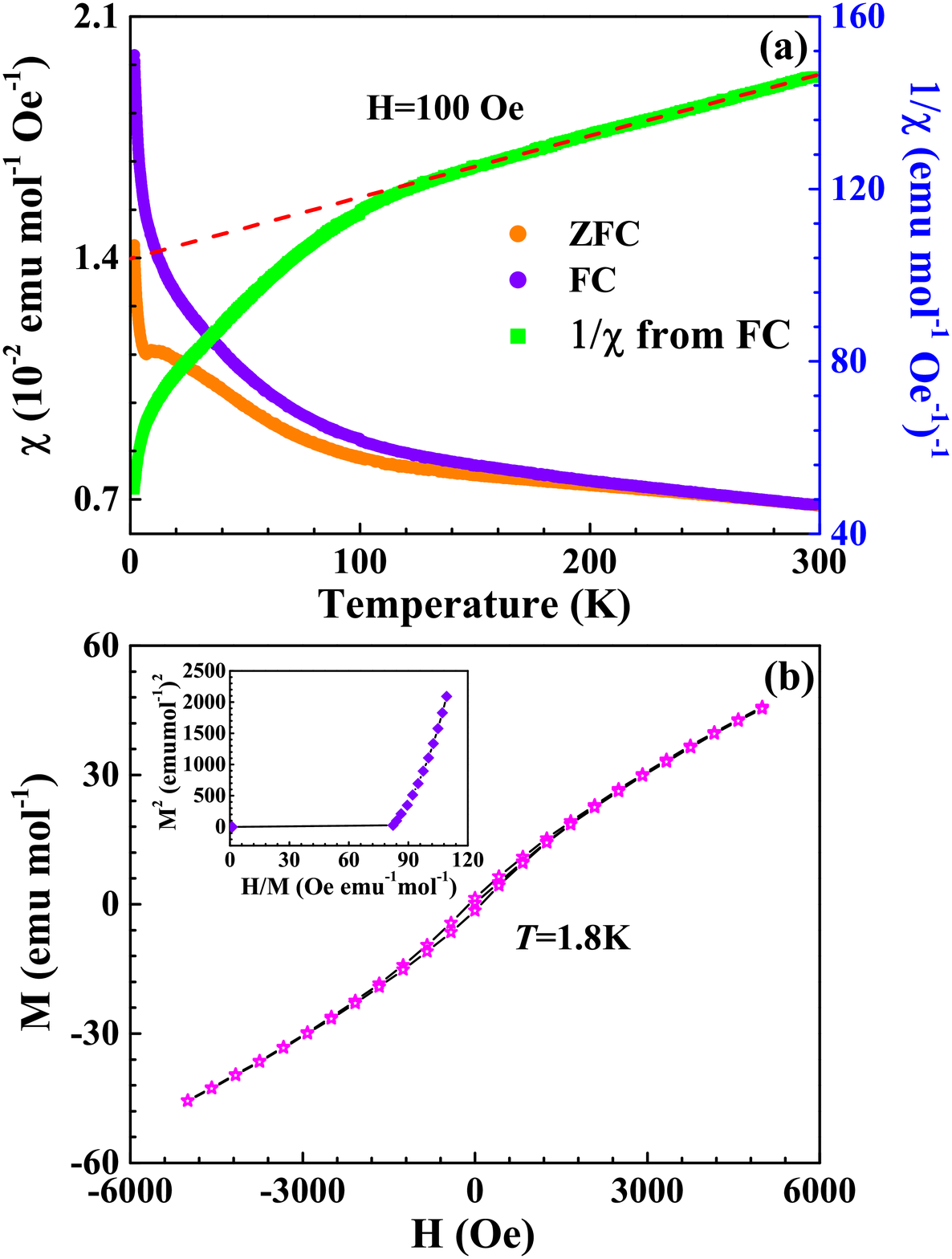}}
\label{modes} \caption{(Color online) (a) Magnetic susceptibility and the inverse susceptibility $vs$ temperature for FeNb$_{0.8}$Ti$_{0.2}$Sb. The dotted line represents the linear extrapolation of the inverse susceptibility $vs$ temperature plots. (b) Magnetization $vs$ magnetic field at 1.8 K. Inset: Arrot plot at 1.8 K of FeNb$_{0.8}$Ti$_{0.2}$Sb.}
\end{figure}

Figure 2(a) shows the temperature dependence of the zero-field-cooled (ZFC) and field-cooled (FC) magnetic susceptibility curves with the external field of 100 Oe together with the inverse magnetic susceptibility ($\chi^{-1} vs$) data in the FC run of FeNb$_{0.8}$Ti$_{0.2}$Sb between 1.8 K and 300 K. Both the ZFC and FC magnetic susceptibilities increase with decreasing temperature, exhibiting a sharp increase at lower temperatures below 6 K. A peak in ZFC is observed around 10 K and irreversibility in ZFC and FC curves occurs below 200 K. A divergence in ZFC and FC along with a peak in the ZFC curves has been reported in the systems\,\cite{vas,ber,all1,rub,all2} that possess mixed exchange interactions, such as spin glass, superparamagnetic or magnetic clusters. Taking other Heusler alloys for reference, the present results suggest the sample is magnetically disordered. The broad maximum in ZFC curve suggests the presence of distribution of magnetic clusters/defects\,\cite{vas}. On the other hand superparamagnetism could also be taken into account\,\cite{vas}. With decreasing temperature, the inverse magnetic susceptibility follows a Curie-Weiss law above 135 K, indicating a paramagnetic behavior. However, it deviates markedly from the Curie-Weiss law below 135 K. Curie-Weiss law has the formula: $\chi(T)=\chi_{0} + C/(T-T_{C})$, where $C$=$N_{A}\mu_{eff}^2/(3k_{B})$, $N_{A}$ is Avogadro's number, $\mu_{eff}$ is the effective moment, $\mu_{B}$ is the Bohr magneton, and $T_{C}$ is the Curie-Weiss temperature. A least-squares fit of the inverse magnetic susceptibility from 135 K to 300 K is shown in Fig.2. The excellent fitting indicates the onset of weak antiferromagnetism below 135 K. The antiferromagnetism is probably a result of atomic disorder\,\cite{sle1,lue,sle2}.

In order to identify the magnetic phase at lower temperatures below 10 K, we investigate the magnetization ($M$) $vs$ magnetic field ($H$) at 1.8 K shown in Fig. 2(b). Superparamagnetism is a form of magnetism which appears in small ferromagnetic or ferrimagnetic nanoparticles. In the absence of an external magnetic field, when the time used to measure the magnetization of the nanoparticles is much longer than the N$\acute{e}$el relaxation time, their magnetization appears to be in average zero. While ferromagnetism is a form of magnetism which could exhibit spontaneous magnetization: a net magnetic moment in the absence of an external magnetic field. As shown in Fig. 3, a so small amount of hysteresis exists at 1.8 K. This is a good indication of the presence of either superparamagnetism or weak ferromagnetism. Because a small hysteresis will also occur in superparamagnetism below the blocking temperature\,\cite{feng,vas}. Therefore, a further investigation of magnetic property is needed. An Arrot plot ($M^{2} vs H/M$) of the $M(H)$ data for $H\leqq$5 kOe at 1.8 K is presented in the inset of Fig. 2(b). The Arrot plots are not linear and the slope is positive, further confirming the superparamagnetism and ruling out the possibility of ferromagnetism\,\cite{vas}. The occurrence of hysteresis is due to freezing of the superparamagnetism below 10 K\,\cite{vas}. The presence of antiferromagnetic state and superparamagnetic clusters below 135 K and 10 K, respectively, in the sample thus can be explained by the existence of atomic disorder. We must emphasize that the exact identification of heusler alloys remains unsettled\,\cite{vas,ber,all1,rub,all2,sle1,lue,sle2,feng}. The magnetic characterization of the sample is not the focus of this article. We concentrate more on the thermoelectric properties and their origins.

\subsection{Specific heat}

\begin{figure}[tbp]
\centerline{\includegraphics[width=0.48\textwidth]{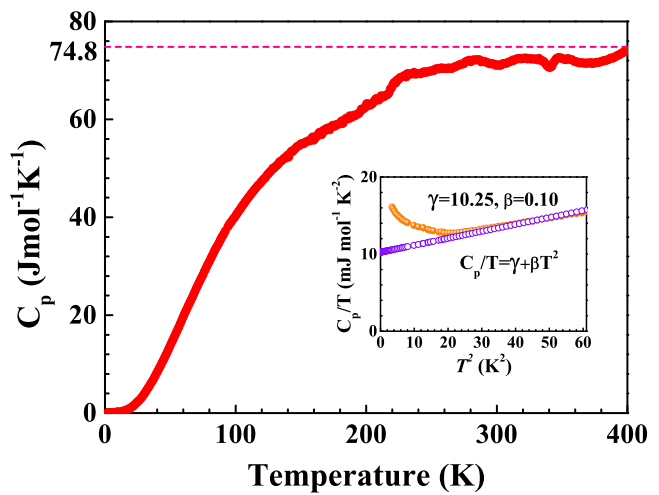}}
\caption{(Color online) Temperature dependence of specific heat of FeNb$_{0.8}$Ti$_{0.2}$Sb. The inset presents the low-temperature data as a $C_P/T$ vs $T^2$ function. The dotted line is the least-squares fit according to the equation above.}
\end{figure}

\begin{figure}[tbp]
\centerline{\includegraphics[width=0.48\textwidth]{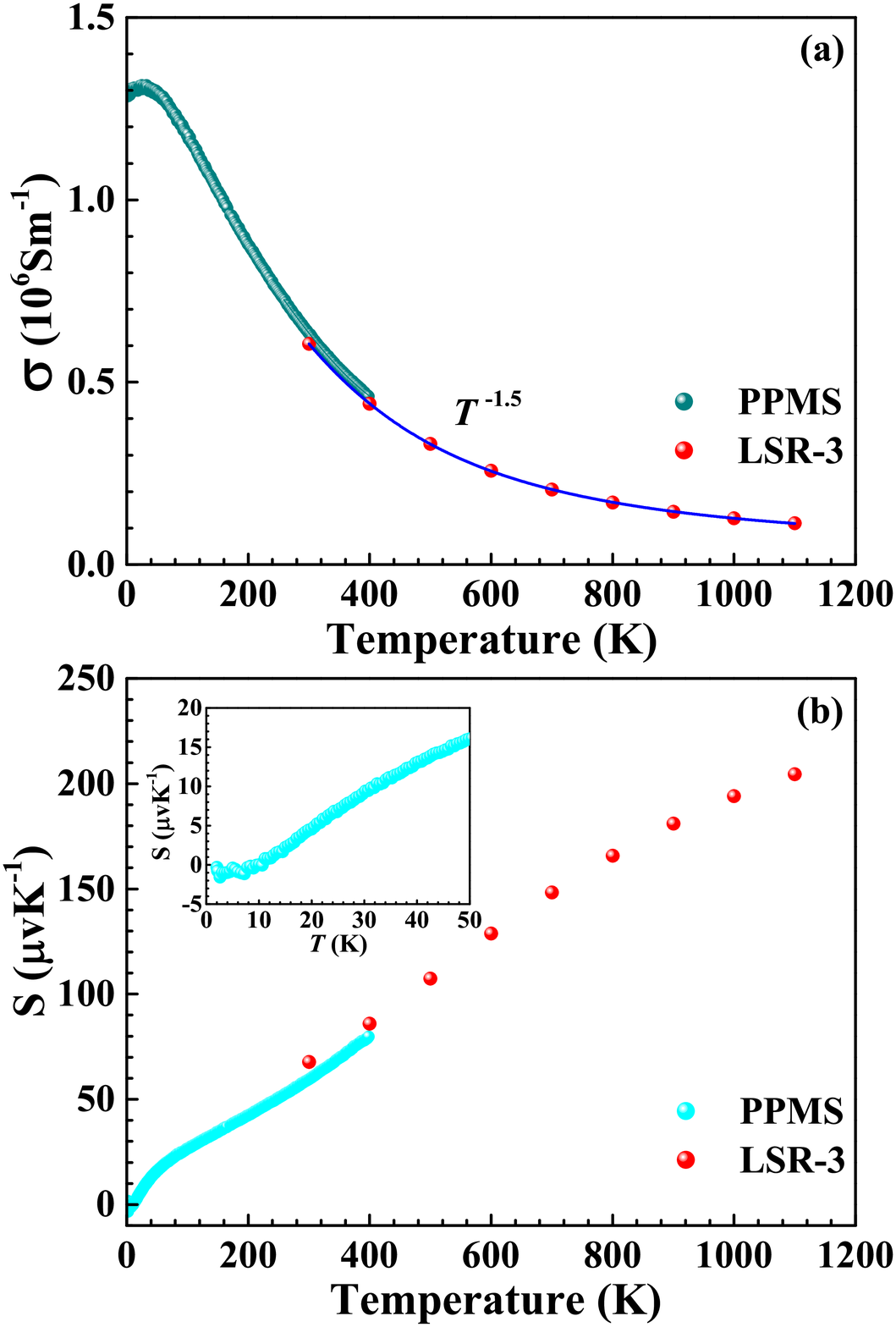}}
\caption{(Color online) Temperature dependence of (a) electrical conductivity and (b) Seebeck coefficient of FeNb$_{0.8}$Ti$_{0.2}$Sb. The inset of (b) is the temperature dependence of Seebeck coefficient below 50 K. The high-temperature data taken from a LSR-3 system are shown in red for comparison.}
\end{figure}

\begin{figure}[tbp]
\centerline{\includegraphics[width=0.48\textwidth]{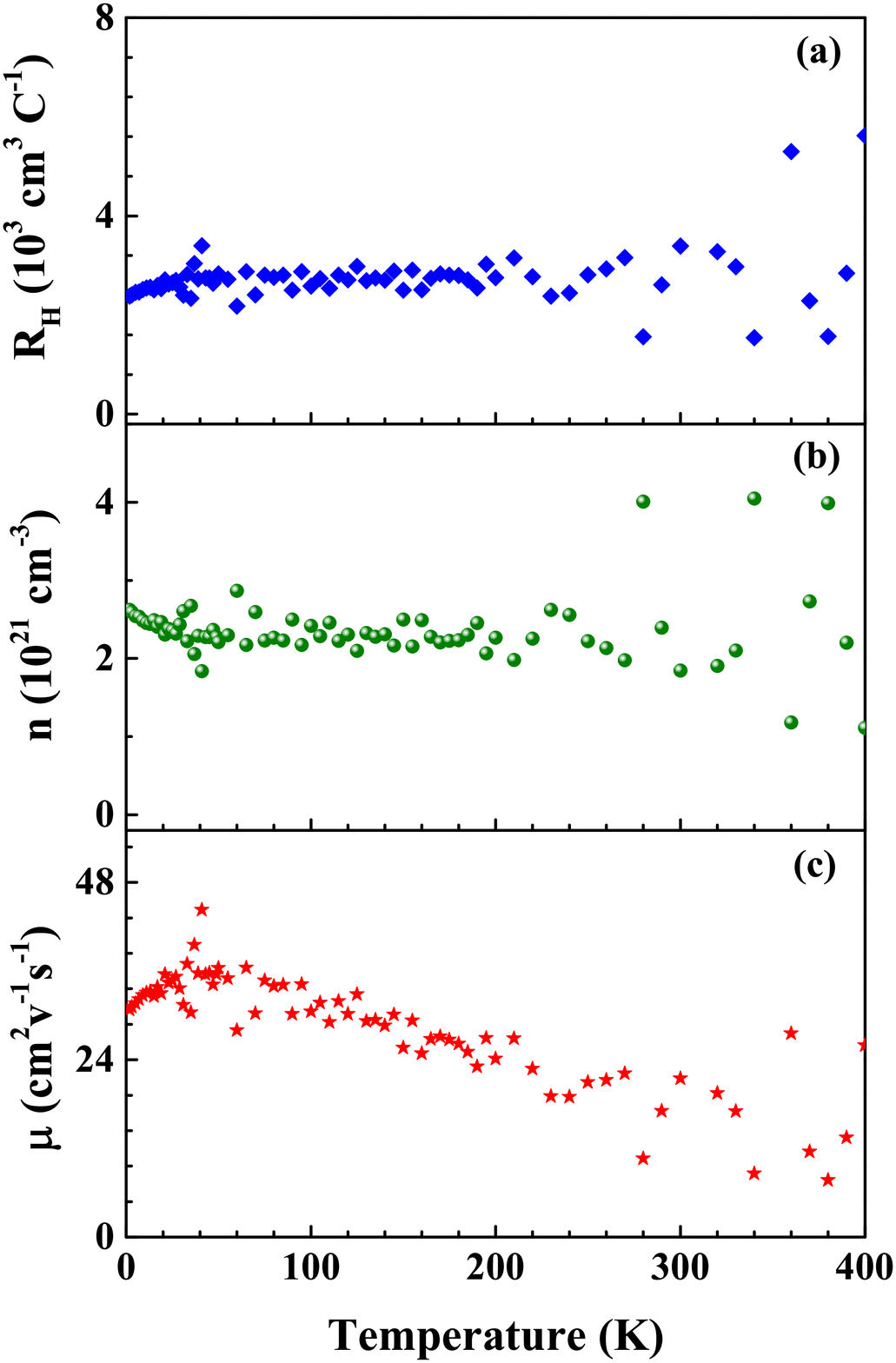}}
\caption{(Color online) Temperature dependence of (a) Hall coefficient, (b) carrier concentration and (c) carrier mobility of FeNb$_{0.8}$Ti$_{0.2}$Sb.}
\end{figure}

Figure 3 presents the temperature dependence of specific heat $C_{p}$ of FeNb$_{0.8}$Ti$_{0.2}$Sb. The specific heat curve has a typical sigmoid like shape and approaches a value expected from Dulong-Petit law, $C_{p}$ = 3$nR$ = 74.8 J mol$^{-1}$ K$^{-1}$, where $n$ is the number of atoms per molecule and $R$ is the gas constant. At very low temperatures, the specific heat in FeNb$_{0.8}$Ti$_{0.2}$Sb gradually diminishes to zero. The inset in Fig. 3 shows the low temperature dependence of specific heat presented as $C_{P}/T$ $vs$ $T^{2}$ from 5 K to 10 K. It can be well described by the formula\,\cite{gof} \vspace{2mm}\newline \centerline { $C_{p}=\gamma T+\beta T^{3}$,} \vspace{1mm}\newline where $\gamma T$ and $\beta T^{3}$ are the electron and phonon contribution to the total specific heat, respectively. As a result, the coefficients $\gamma$ is 10.25 mJ mol$^{-1}$ K$^{-2}$, and $\beta$ is 0.10 mJ mol$^{-1}$ K$^{-4}$. From the value of $\beta$ one can estimate the Debye temperature $\Theta_D$=$(12R\pi^4n/5\beta)^{1/3}$ to be about 388 K. An abnormal upturn seen at low temperature is similar to that observed in several systems, including the new iron-based superconductors and other heusler materials\,\cite{kim,gof2,dor}. For FeNb$_{0.8}$Ti$_{0.2}$Sb, the phenomenon maybe origin from the magnetic clusters arising from the atomic disorder.

It's noteworthy that there has been a theoretical estimate the thermal conductivity using Debye theory and could reveal a relation between thermal conductivity and specific heat, which is given by $\kappa=\frac{1}{3}C\nu l$, where $C$ is the specific heat per volume, $\nu$ is the average phonon velocity, and $l$ is the phonon mean free path. At the very low temperatures, the low specific heat indicates the low thermal conductivity. With increasing temperature, the specific heat increases fast and approaches to a constant value. Therefore the thermal conductivity will increase rapidly and also reach a maximum. At higher temperatures, with the enhancement of the phonon-phonon scattering, the average phonon velocity and phonon mean free path are limited significantly and the thermal conductivity reduces largely.

\subsection{Electrical transport properties}

Figure 4 illustrates the temperature dependence of (a) electrical conductivity and (b) Seebeck coefficient of FeNb$_{0.8}$Ti$_{0.2}$Sb. The high-temperature data taken from LSR-3 system are drawn in red color for comparison. As shown in the figures, the low and high temperature electrical transport properties measured in the different methods are consistent with each other. In addition, the low and high temperature data converge at room temperature. As temperature is increased, the electrical conductivity decreases rapidly in the range of 10$^6$ $S m^{-1}$, following the degenerate semiconducting behavior\,\cite{cgf2}. This implies that the electrical conductivity will follow a temperature dependence of $T^{-1.5}$ from the Debye temperature (388 K) to the intrinsic excitation temperature\,\cite{cgf3}, which agrees well with the high-temperature experiment data (388 K - 1100 K). It means that acoustic phonon scattering dominates charge transport\,\cite{shi}, which is consistent with the specific heat measurement. There is an upturn at the low temperatures (below 30 K). The anomalous temperature dependence of electrical conductivity maybe due to the magnetic clusters arising from the atomic disorder\,\cite{dor}.

The values of Seebeck coefficient are negative below 10 K and keep positive from 10 K to 1100 K. As temperature is increased, the Seebeck coefficient increases rapidly and approaches 80 $\mu$vK$^{-1}$ in the vicinity of 400 K and approaches a maximum of 205 $\mu$vK$^{-1}$ at 1100 K, which is typical behavior for degenerate semiconductors\,\cite{cgf2}. Thus it can be predicted that the Seebeck coefficient will linearly increase with increasing temperature before the intrinsic excitation which is in accordance with the high-temperature data.

For FeNb$_{0.8}$Ti$_{0.2}$Sb, the carrier concentration is one of the most important physical parameters for thermoelectric performance. The electrical conductivity is related to carrier concentration through carrier mobility ($\mu$): $\sigma=ne\mu$, where $e$ is the unit charge. The carrier concentration is calculated by $n=1/eR_{H}$, where $e$ is the unit charge and $R_{H}$ is the Hall coefficient\,\cite{fu}. Figure 4 shows the temperature dependence of (a) Hall coefficient, (b) the calculated carrier concentration and (b) carrier mobility of FeNb$_{0.8}$Ti$_{0.2}$Sb from 1.8 K to 400 K, respectively. The carrier concentration is rather constant and almost independent of temperature, about 10$^{21}$ $cm^{-3}$ below 400 K which is in the optimal value range\,\cite{sny,fu}. The carrier mobility decreases slightly with temperature and becomes 25 $cm^2v^{-1}s^{-1}$ at 400 K, which is consistent with the electrical conductivity. The Hall coefficient values are positive and keep stable over the whole low temperature range, indicating the majority of the charge carriers are holes and there is only a single type of carrier which will benefit the Seebeck coefficient. For degenerate semiconductors the Seebeck coefficient is given by\,\cite{sak} \vspace{2mm}\newline\centerline {$S = \frac{8\pi^2\kappa^2_BT}{3eh^2}m^*(\frac{\pi}{3n})^{2/3}$,} \vspace{1mm}\newline where $m^{*}$ is the effective mass of the carrier. Thus the large Seebeck coefficient is due to the optimal carrier concentration and hole carriers which have larger effective mass than electrons. Below 10 K, the values of Seebeck coefficient are below zero, implying the effective mass is negative. This means the band curves downwards away from a maximum. Taking into account the magnetic properties of FeNb$_{0.8}$Ti$_{0.2}$Sb, the magnetic clusters arising from the atomic disorder in the sample maybe contribute to the phenomenon.

\subsection{Thermal transport properties}

\begin{figure}[tbp]
\centerline{\includegraphics[width=0.48\textwidth]{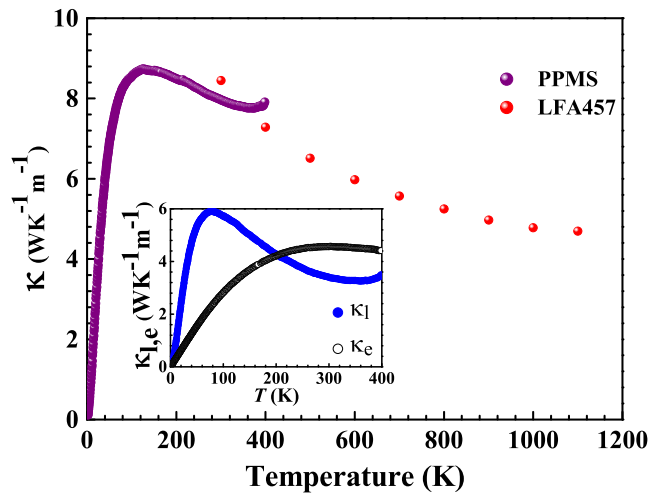}}
\caption{(Color online) Temperature dependence of the thermal conductivity of FeNb$_{0.8}$Ti$_{0.2}$Sb, as well as high-temperature data (in red) estimated by a laser flash method on a Netzsch LFA457 instrument with a Pyroceram standard. The inset is the temperature dependence of the lattice and electron components of the thermal conductivity of FeNb$_{0.8}$Ti$_{0.2}$Sb.}
\end{figure}

Figure 6 presents the temperature dependence of the thermal conductivity of FeNb$_{0.8}$Ti$_{0.2}$Sb, as well as high-temperature data estimated by a laser flash method on Netzsch LFA457 instrument with a Pyroceram standard. The data from low temperature agree well with the high-temperature data measured in different methods. As temperature is increased, the thermal conductivity increases rapidly and reaches a maximum (approximately 8.7 WK$^{-1}$$m^{-1}$) around 126 K, and then declines gradually, which is in accordance with the estimation from specific heat. The scattering mechanism could give further explanation of the shape of the observed thermal conductivity curve as follows: the thermal conductivity is typically limited by normal three-phonon scattering, umklapp scattering, impurity scattering and boundary scattering\,\cite{cal}. At very low temperatures, boundary scattering predominates the scattering mechanism and the thermal conductivity is small. As temperature is increased, the impurity scattering becomes important because it becomes easier to create higher frequency phonons which are scattered efficiently by point impurities and the thermal conductivity reaches a maximum and then declines. As temperature is increased further, normal three-phonon scattering and umklapp scattering gradually come to dominate. At higher temperatures, all phonon scattering occupies the scattering mechanism.

The inset shows the temperature dependence of the lattice and electron components of the thermal conductivity of FeNb$_{0.8}$Ti$_{0.2}$Sb. The lattice thermal conductivity was obtained by subtracting the electron component from the total thermal conductivity. The electronic thermal conductivity was calculated via $\kappa_{e}$=$L\sigma T$=$Lne\mu T$, where $L$ is Lorenz number and can be calculated using the single parabolic band (SPB) model with reasonable approximation\,\cite{xie}. The temperature dependence of lattice thermal conductivity is similar to that of the total thermal conductivity and the maximum is 5.9 WK$^{-1}$$m^{-1}$ around 77 K. In previous work, it's found the carrier mean path in the $p$-type FeNbSb is comparable to the lattice parameter, indicating that the carrier mobility of this system almost reaches the loffe-Regel limit\,\cite{Gurvitch}, which means that the carrier scattering has reached the highest limit and introducing more phonon scattering centers will not impair the power factor while largely suppress the lattice thermal conductivity\,\cite{cgf2}. For semiconductors, the electronic thermal conductivity is much less than the lattice thermal conductivity, while the electron contribution to the total thermal conductivity of FeNb$_{0.8}$Ti$_{0.2}$Sb is not negligible. As shown in Fig. 6, the electronic thermal conductivity increases with temperature and becomes higher than the lattice thermal conductivity from 200 K.

To minimize the lattice thermal conductivity, disorder within the unit cell\,\cite{sale}, superlattices\,\cite{venk}, complex unit cells\,\cite{chun}, nanostructures\,\cite{tjz2} have been widely used in the thermoelectric materials over the past several years. For FeNb$_{0.8}$Ti$_{0.2}$Sb, the electronic thermal conductivity is comparable to or even higher than the lattice thermal conductivity above 200 K. This means the lattice thermal conductivity is largely suppressed and the thermal conductivity is mainly determined by the electronic thermal conductivity.

\subsection{Figure of merit zT}

\begin{figure}[tbp]
\centerline{\includegraphics[width=0.48\textwidth]{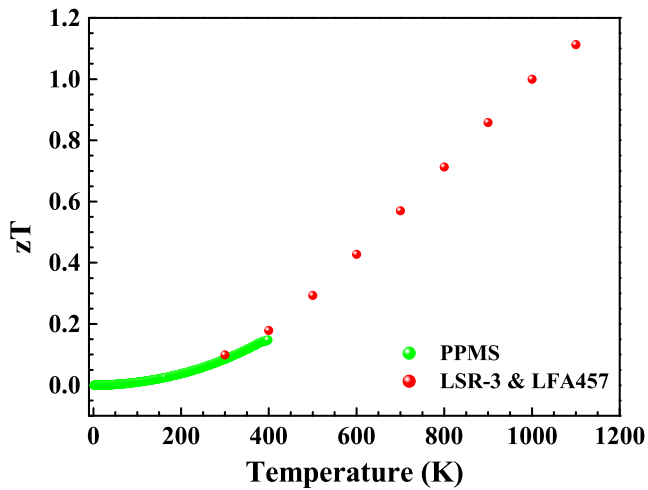}}
\caption{(Color online) Temperature dependence of $zT$ of FeNb$_{0.8}$Ti$_{0.2}$Sb. The data at high temperatures (in red) are taken from a LSR-3 system and a Netzsch LFA457 instrument.}
\end{figure}

Figure 7 shows the temperature dependence of $zT$ of FeNb$_{0.8}$Ti$_{0.2}$Sb in the temperature range of 1.8 - 400 K, together with the high-temperature data measured in different method for comparison. As the intensive result of the electrical conductivity, Seebeck coefficient and thermal conductivity, the $zT$ exhibits a pronounced rise with temperature. Afterwards, it keeps a continuous increase to 0.14 around 400 K and reaches a maximum of 1.1 at 1100 K. Compared with other high temperature thermoelectric materials, FeNb$_{0.8}$Ti$_{0.2}$Sb exhibits excellent thermoelectric performance for power generation. The $zT$ exceeds the industry benchmarks set by $p$-type silicon-germanium high temperature alloys\,\cite{use}. Even more, it's better than the optimized $n$-type (Hf, Zr)NiSn half-Heusler compound (the maximum $zT$ is 1.0 at 1000 K)\,\cite{tjz1,dow,che}. The values of $zT$ at low temperatures join well with that of high temperatures and converge with the high-temperature data around room temperature. The trend demonstrated in the case of low temperature is extended to high temperature. This means the large Seebeck coefficient and the moderate electrical conductivity as well the relatively low thermal conductivity at low temperatures, which result from an optimal and temperature-independent carrier concentration and high content of Ti doping, will account for that of high temperatures. It's known that power factor is essentially determined by the electronic properties, and the thermal conductivity of FeNb$_{0.8}$Ti$_{0.2}$Sb is also mainly dominated by electron component. Based on the above considertaion, the $zT$ of FeNb$_{0.8}$Ti$_{0.2}$Sb is mainly governed by the electronic properties.

\section{CONCLUSIONS}

In conclusion, we have performed electrical and thermal transport measurements on FeNb$_{0.8}$Ti$_{0.2}$Sb at low temperatures in order to elucidate the physical origins of the high thermoelectric performance avoiding the influence of thermal fluctuations. The low-temperature trend of the electrical conductivity, Seebeck coefficient and thermal conductivity are extended to high temperature. The optimized power factor mostly results from the optimal and almost temperature-independent carrier concentration. Meanwhile, a single type of hole carrier benefits the Seebeck coefficient as well. The lattice thermal conductivity is largely suppressed and the total thermal conductivity is mainly determined by the electronic thermal conductivity. Consequently, the $zT$ of FeNb$_{0.8}$Ti$_{0.2}$Sb is mainly governed by the electronic properties. As a result, the $zT$ exhibits a pronounced rise from the low to high temperatures and approaches a maximum of 1.1 at 1100 K exceeding that of state-of-the-art thermoelectric materials. These findings highlight investigating low-temperature physical properties of thermoelectric materials can help to have a thorough knowledge of their behavior at high temperatures.

\vspace{8mm}
\begin{acknowledgments}
We thank Yanping Yang for help in the Energy Dispersive Spectrometer measurement. The sample preparation was supported by Natural Science Foundation  of China (No. 11574267).
\end{acknowledgments}

\end{document}